\documentclass[aps,prb,footinbib,reprint,superscriptaddress]{revtex4-2}

\usepackage{graphicx}
\graphicspath{ {images/}} 
\usepackage{dcolumn}
\usepackage{bm}
\usepackage[caption=false]{subfig}
\usepackage{amsmath}
\usepackage{amsfonts}
\usepackage{footmisc}

\begin{document}

	\title{Topologically Protected Edge States in Triangular Lattices}
	
	\author{Robert J. Davis}
	\email{rjdavis@ucsd.edu}
	\affiliation{Electrical and Computer Engineering Department, University of California, San Diego, La Jolla, California 92093, USA}
	\author{Yun Zhou}
	\affiliation{Mechanical and Aerospace Engineering, University of California, San Diego, La Jolla, California 92093, USA}
	\author{Dia'aaldin J. Bisharat}
	\affiliation{Advanced Science Research Center, The City University of New York, New York, NY 10031, USA}
	\author{Prabhakar R. Bandaru}
	\affiliation{Mechanical and Aerospace Engineering, University of California, San Diego, La Jolla, California 92093, USA}
	\author{Daniel F. Sievenpiper}
	\email{dsievenpiper@ucsd.edu}
	\affiliation{Electrical and Computer Engineering Department, University of California, San Diego, La Jolla, California 92093, USA}

	\date{\today}
	
	\begin{abstract}
		We describe the possibility for topologically robust edge states existing on interfaces of triangular lattices which are supported by rotational symmetries that are sensitive to boundary conditions. Such states are trivial from the perspective of Berry curvature, but result instead from an interplay between crystalline symmetries and finite boundary effects. We show such states comprise a distinct topological phase, provided the gauge-dependent symmetries are maintained. Such a model describes a number of recent bosonic experimental demonstrations on triangular lattices, the physics for which has thus far eluded explanation. 
	\end{abstract}
	
	\maketitle
	
	
	\section{Introduction}
	Recent advances in topological physics have revealed a wide class of nontrivial phases that can exist in condensed matter systems, each relying upon maintaining or breaking various symmetries \cite{hasan_colloquium:_2010}. These studies began with the quantum Hall effect \cite{thouless_quantized_1982} and related time reversal symmetry ($\mathcal{T}$) broken systems, but later were generalized to spin-based platforms that preserve $\mathcal{T}$ symmetry \cite{kane_quantum_2005}. Still more recently, many experimental demonstrations have explored the use of various crystalline symmetries to create topological insulators (TIs) \cite{fu_topological_2011}, owing to their simplicity of implementation in bosonic systems.  Such crystalline symmetry-protected phases have been demonstrated for systems in square lattices \cite{liu_novel_2017}  and Kagome crystals \cite{ezawa_higher-order_2018,ni_observation_2019,li_higher-order_2020}, and can be well characterized by their various rotation eigenvalues at high symmetry points in the Brillouin zone (BZ). In each case, such phases require a minimum of orbital sites within a unit cell to define the given rotational symmetry (e.g., 4 four the square lattice, 3 for the Kagome, etc.); as such non-primitive cells are required for each. 
	
	These crystalline phases stand in contrast to the earlier Chern \cite{thouless_quantized_1982}, spin \cite{kane_quantum_2005}, and valley \cite{xiao_valley-contrasting_2007} phases, which are defined by topological invariants computed in reciprocal space, as they instead involve information of the real space defined configuration of the system. The earliest example of Ref. \cite{fu_topological_2011} showed how point group symmetries can induce a phase possessing gapless surface states, which are otherwise trivial in the framework of earlier topological classification systems \cite{altland_nonstandard_1997, ryu_topological_2010}. More recent studies into the influence of crystalline symmetries has yielded a plethora of new phenomena, including higher order topological insulators \cite{schindler_higher-order_2018} and surface rotation anomalies \cite{fang_new_2019,fan_discovery_2021}. These demonstrations have been recently unified under more general notions of symmetries based on point and space groups, commonly referred to as symmetry indicators \cite{fang_bulk_2012,slager_space_2013,kruthoff_topological_2017,po_symmetry-based_2017,benalcazar_quantization_2019}, which rely on information of the real space configuration as well as knowledge of the wavefunctions  at various high symmetry point in the BZ. These techniques reveal a broad class of topologically nontrivial structures in real material systems, which have been efficiently tabulated \cite{tang_comprehensive_2019,tang_efficient_2019}.
	
	Such phases are frequently referred to as topological, insofar as they define a global property of the band structure and can be described by an invariant that changes discretely \cite{fang_bulk_2012}. This naturally leads to a gauge-dependence for the various topological invariants that characterize them, in sharp relief to those in other systems. This can be seen even in the 1D Su-Schrieffer Heeger model, where the Zak phase depends on the choice of unit cell, though the difference of two such choices is unique  \cite{atala_direct_2013}. Throughout the paper we will refer to "topologically protected" to include such gauge-dependent systems, as well as obstructed atomic orbital states \cite{benalcazar_quantization_2019}. 
	
	Recently, a number of physical systems in photonics \cite{yang_evolution_2021, bisharat_robust_2021} and phononics \cite{zhou_-chip_2021} have demonstrated a form of unidirectional propagation for bosons on triangular lattices within a defect line. Such platforms have zero Berry curvature \cite{liu_novel_2017,liu_topological_2018}, and as such appear trivial from the spin and valley perspectives. We will show via a tight binding model that these systems can in fact be described by a non-trivial topology based on a specific flavor of symmetry indicator that focuses on rotational symmetries \cite{benalcazar_quantization_2019}. Specifically, in this letter we demonstrate that a triangular lattice with $C_{3v}$-symmetric hopping terms can lead to topologically protected edge states. 
	
	\section{Tight Binding Model and Reciprocal Space Characteristics}
	
	\begin{figure}
		\centering
		\includegraphics[width = \linewidth]{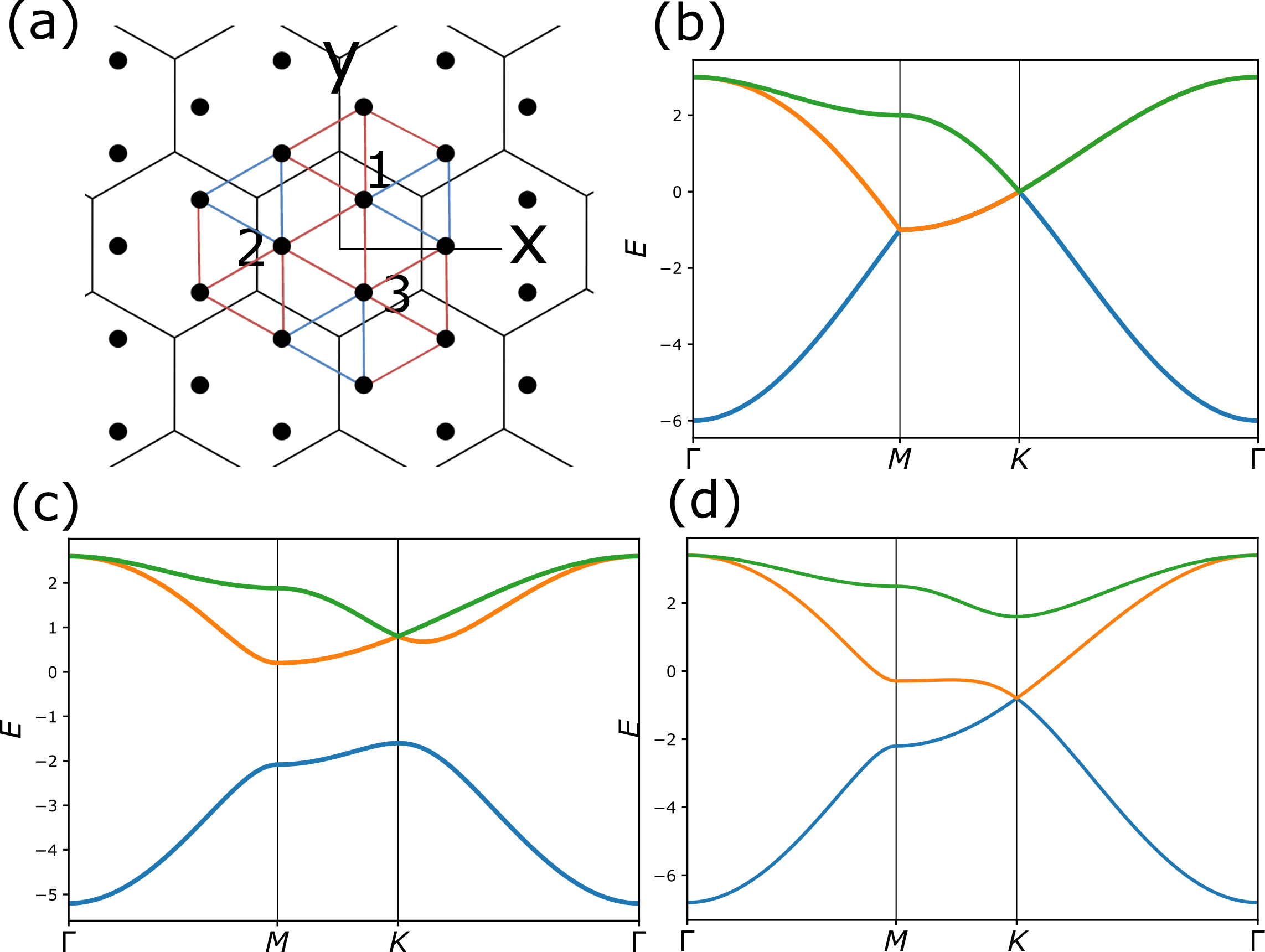}
		\caption{Triangular Lattice in a three-band model. (a) Diagram of unit cell in chosen basis, with $t_+$ bonds shown in blue and $t_-$ bonds in red. (b)-(d) Band structure of the first three bands of the (b) pure triangular lattice with equal hopping $\delta = 0$ ($t_+ = t_-$), (c) nontrivial gapped hopping $\delta < 0$ ($t_+>t_-$), and (d) Dirac-cone hopping $\delta > 0$ ($t_+ < t_-$). }
		\label{fig1}
	\end{figure}
	\begin{figure*}
		\centering
		\includegraphics[width = \linewidth]{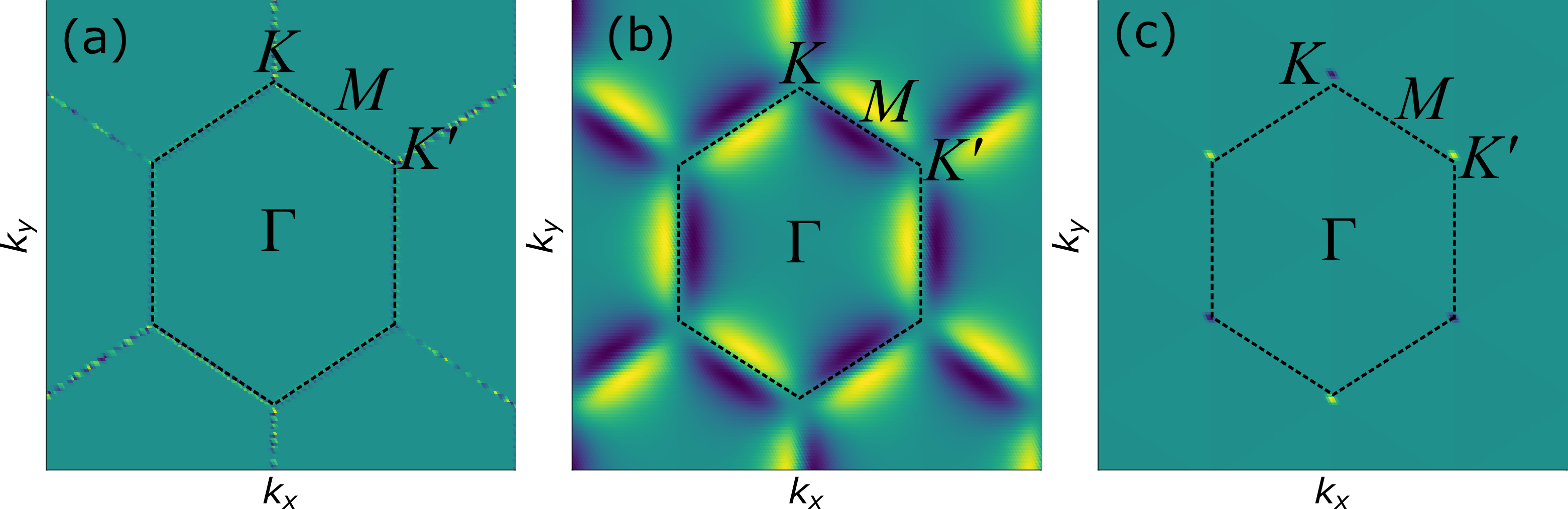}
		\caption{Berry curvature distributions for the (a) pure triangular, $\delta = 0$, (b) nontrivial triangular, with a typical gapped value $\delta = 0.4$, and (c) Dirac cone-like, $\delta = -0.4$ cases. In (a) there is rapid fluctuations along the degeneracies at boundary of the reduced BZ (noted by the black dotted line) which average to zero. For (c), a small staggered on-site potential of $10^{-3}$ was added so the sign of the singularities at $K/K'$ were uniquely defined.}
		\label{fig2}
	\end{figure*}
	
	We adopt a Hamiltonian on a triangular lattice with the hopping texture as shown in Fig. \ref{fig1}(a), given generically as 
	\begin{equation}
		H = -\sum_{\langle ij\rangle} t_{\pm}c^\dagger_i c_j +\text{H.c}
	\end{equation}
	Here, $\langle ij\rangle$ denotes nearest neighbor hopping from site $i$ to site $j$, and $t_{\pm}\equiv 1\pm\delta$ describes the texture of the hopping terms. We will initially set the onsite potential to zero and limit the analysis to the region of $-1\le \delta \le 1$. We adopt a three-site basis as illustrated in Fig. \ref{fig1}(a) with kernel of the Bloch Hamiltonian
	\begin{equation}
		H(\delta) = \begin{pmatrix}
			0 & h_{12}^* & h_{13}^* \\
			h_{12} & 0 & h_{23}^* \\
			h_{13} & h_{23} & 0
		\end{pmatrix}.\label{H}
	\end{equation}
	where	
	\begin{align*}
		h_{12} &= t_- + t_+ e^{ik_x} +  t_- e^{i(k_x/2 +\sqrt{3}/2 k_y)}\\
		h_{13} &= t_-+ t_+  e^{i(k_x/2 +\sqrt{3}/2 k_y)} +t_- e^{-i(k_x/2 -\sqrt{3}/2 k_y)}\\
		h_{23}&= t_- + t_+ e^{-i(k_x/2 -\sqrt{3}/2 k_y)}+ t_- e^{-i k_x}
	\end{align*}
	
	The model obeys time reversal symmetry, and falls into class AI of the Altland-Zirnbauer classification \cite{altland_nonstandard_1997,ryu_topological_2010}. Note that the form of $H$ is similar to Kagome lattices \cite{ezawa_higher-order_2018}, but here each site has 6 nearest neighbors, rather than 4. This has an important consequence in that the low energy band structure is degenerate at all $\mathbf{k}$ values along the $M-K$ boundary for $\delta = 0$, rather than the Dirac degeneracy seen in Kagome models. These extra band degeneracies are not protected by rotational symmetry, as the little group of the wave vector at the $M$ point for the lattice ($C_{2v}$) does not permit any 2D irreducible representations \cite{malterre_symmetry_2011} (See Appendix \ref{group} for more details). Nevertheless, this difference from Kagome or honeycomb models manifests in the symmetry properties of the Berry phase and how they determine the existence of edge states. 
	
	In the ideal triangular lattice with unity potential, $\delta = 0$, we have the degenerate band structure seen in Fig. \ref{fig1}(b). If we modify the hopping such that $\delta > 0 $, a band gap is opened for the lowest band, as shown in Fig. \ref{fig1}(c). For the opposite case of $\delta < 0$, a $C_{3v}$-protected Dirac cone is found, shown in Fig. \ref{fig1}(d). In the following sections we analyze these three cases individually, and show how the latter, gapped case posses an interesting question not readily solved with reciprocal space techniques. 
	
	\subsection{Ideal $\delta = 0$ Case}
	The ideal triangular lattice under a tight binding (TB) formalism, Fig. \ref{fig1}(b) does not have a bandgap, and therefore cannot demonstrate any edge states independent from bulk states. However, as we will further detail in Sec. \ref{experiments}, such systems do in fact posses a fundamental bandgap within bosonic systems. 
	
	Fig. \ref{fig2}(a) shows the Berry curvature distribution for Eq. \ref{H} under this case. From the combination of $\mathcal{T}$ and inversion symmetry $\mathcal{I}$, the curvature is pinned to zero for all values within the Brillouin zone, except those along the points of degeneracy, where the non-Abelian form of the curvature must be used to determine the values. Here we employ the Abelian form, and as a result we observe rapid numerical fluctuations along the BZ edges that average to zero \cite{liu_novel_2017}. 
	
	\subsection{Dirac $\delta<0$ Case}
\begin{figure*}
		\centering
		\includegraphics[width = \linewidth]{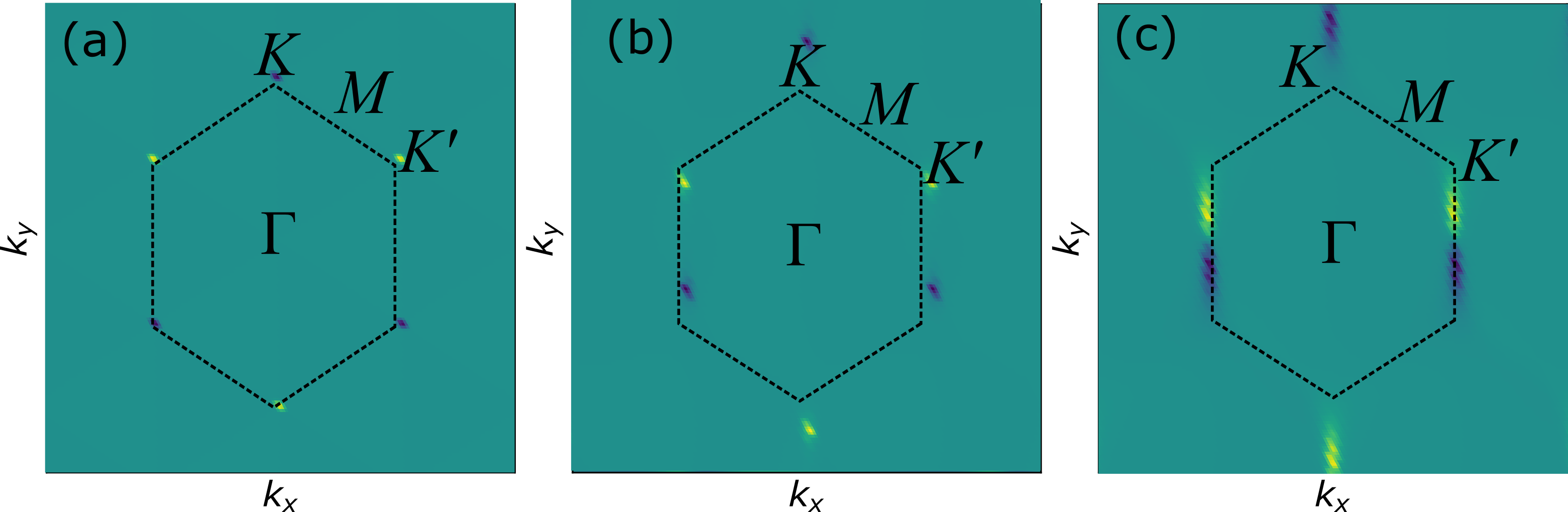}
		\caption{Berry curvature distribution for the Dirac-like case of $\delta <0$ with variable staggered onsite-potential $d$ for (a) $d = 10^{-4}$, (b) $d = 0.5$, (c) $d = 1.5$. In each figure a representative value of $\delta = -0.4$ is used. We see that as the staggered potential is increased, the distribution becomes asymmetric, with $K/K'$ singularities becoming poorly defined.}
		\label{fig:berry}
	\end{figure*}

	The Dirac case, Fig. \ref{fig1}(d) is reminiscent of the valley Hall physics of graphene \cite{xiao_valley-contrasting_2007}, but here a difference arises in how a gap can be introduced. Namely, if a staggered onsite potential is applied, a gap will appear near $K/K'$, but in doing so the point group is lowered to $C_s$, rather than the $C_{3v}$ of a valley-like model. This causes the location of the Berry curvature singularity to shift from $K/K'$, deteriorating any resulting edge states as the "valleys" are no longer at $\mathcal{T}$-linked locations in the BZ. To show this in reciprocal space, Fig. \ref{fig:berry} shows the evolution of the Berry curvature as the alternating on-site potential is increased. Here the definition for "alternating" is 0, $+d$, $-d$ for sites 1, 2, and 3 of the unit cell as labeled in Fig. 1(a).
	
	It can be observed that as soon as the on-site potential $d$ is non-zero, the Dirac cone is gapped and the singularities form well-defined peaks at the $K/K'$ valleys (Fig. \ref{fig:berry}(a)). However, as $d$ is increased, Fig. \ref{fig:berry}(b)-(d), we see the two peaks drift from the valleys, destroying the valley-like behavior and the valley-projected Hamiltonian will not have a well-defined valley Chern number \cite{li_marginality_2010}. 
	
	\subsection{Gapped $\delta>0$ Case}
	
	The gapped case, with $\delta > 0$, is different. It is clear that doing so reduces the point group from $C_{6v}$ down to $C_{3v}$, which permits nonzero Berry curvature via breaking of inversion symmetry (Fig. \ref{fig2}(b)). Unlike effective Hamiltonians defined near the $K/K'$ point in valley models, however, the degeneracy being lifted is along the outer boundaries of the BZ rather than the point degeneracy at $K/K'$, and so the resulting Berry phase accumulates along the $M-K$ path with a 3-fold rotational symmetry, provided the correct gauge is chosen \cite{dobardzic_generalized_2015} (see Appendix \ref{gauge} for details). As such, standard valley-polarized states cannot appear in this case either. As we will see, however, such a situation does indeed give rise to surface states, but of a different nature. 
	
	\section{Symmetry Indicators of Rotational Invariants}
	
	As the Berry curvature of the gapped ($\delta >0$) phase does not reveal the topological properties, we turn instead to the symmetry properties of each band by the behavior of their eigenstates at the high symmetry points (HSPs) when acted on by various rotation operators \cite{benalcazar_quantization_2019}. Importantly, unlike the gauge-invariant behavior of the Berry curvature, such symmetry behaviors can be influenced by transformations to the real space configuration of the system. More specifically, for a given $n$-fold rotation operator $\hat{r}_n$, we seek the expectation $\langle \hat{r}^u_n(\Pi)\rangle=\langle u(\Pi) | \hat{r}_n | u(\Pi) \rangle$ for an eigenstate $u$ evaluated at the HSP $\Pi$. In the ideal triangular lattice the relevant rotations are $\hat{r}_3$ and $\hat{r}_6$, but within the modified hopping terms (which break $C_{6v}$ symmetry) we will only need $\hat{r}_3$ \cite{fang_bulk_2012}. 
	
	In the chosen basis, the three fold rotation operator $\hat{r}_3$ can be represented as
	
	\begin{equation}
		\hat{r}_3=\begin{pmatrix}
			0 & 0 & 1 \\
			1 & 0 & 0 \\
			0 & 1 & 0
		\end{pmatrix}.
	\end{equation}
	To evaluate the topology, we must calculate $\langle \hat{r}^u_n(\Pi)\rangle$  for each occupied band at certain HSPs, which we here set to the lowest band only (1/3rd filling), as we are concerned with edge states within the first bandgap. From the theory of \cite{fang_bulk_2012} and \cite{benalcazar_quantization_2019}, we can then evaluate the topological invariant associated to this rotation operator, given as a vector of two integers
	\begin{equation}
		\chi^{(3)} = ([K^{(3)}_1],[K^{(3)}_2]),\label{chi}
	\end{equation}
	where $[K^{(3)}_1]$ and $[K^{(3)}_2]$ are given as 
	\begin{equation}
		[K^{(3)}_p] = \#K^{(3)}_p - \#\Gamma^{(3)}_p,\label{eigs}
	\end{equation}
	and $\#\Pi^{(3)}_p$ is the number of occupied bands with eigenvalue $\Pi^{(3)}_p = e^{2\pi i (p-1)/3}, \quad p = 1,2,3$, for the HSPs $\Pi = K,\Gamma$. 
	
	For the sake of generality, we note that to include the degenerate cases of $\delta\le 0$, we may evaluate $\chi^{(3)}$ by determining the eigenvalues of the overlap matrix $S_{jk}(\Pi)\equiv \langle u_j(\Pi) | \hat{r}_n | u_k(\Pi) \rangle$, where $j,k=1,2,3$ are the band indices. However, this will naturally give the $n$-band manifold's invariant, which is not of interest here (see Appendix \ref{comp} for further details).

	In our case we have simplified the expressions from \cite{benalcazar_quantization_2019} to the case of 3-fold symmetry. In the case of the 6-fold symmetric case of Fig. \ref{fig1}(b) (valid only at $\delta = 0$) the rotational invariant is instead $\chi^{(6)} = ([M^{(2)}_1],[K^{(3)}_1])$, which can be found to be trivial by considering an expanded 6-site basis TB model. Likewise, any symmetry properties for other invariants on this basis are also trivial for the 1/3rd filling case. At the critical point of $\delta > 0$, however, we observe a phase transition where $\chi^{(3)}  = (-1,1)$, indicating a nontrivial topology. We note here that such a phase is topologically equivalent to the $h^{(3)}_{2b}$ primitive generator Hamiltonian from Ref. \cite{benalcazar_quantization_2019}, which possesses an identical $\chi^{(3)}$.
	
	If we rotate the site assignments of the Bloch Hamiltonian Eq. \eqref{H} by $C_2$, or, equivalently, perform a $C_2$ rotation on the Brillouin zone which swaps the $K$ and $K'$ points, the band structure remains identical to that shown in Fig. \eqref{fig1}(d). The difference manifests when considering the symmetry indicator: in this new rotated basis, we find $\chi^{(3)}  = (-1,0)$ from the differing phase of the $K'$ point. This new Hamiltonian is topologically equivalent to the $h^{(3)}_{2c}$ primitive generator of Ref. \cite{benalcazar_quantization_2019}. This implies that a geometrical rotation can result in differing topological phases, which is the mechanism that several recent studies \cite{bisharat_robust_2021, yang_evolution_2021,zhou_-chip_2021,wen_designing_2022} have exploited to realize unidirectional modes in bosonic platforms, which will be discussed in Section \ref{experiments}.
	
	\section{Edge States on Finite Lattices}\label{sec:fin}
	
	\begin{figure}
		\centering
		\includegraphics[width = \linewidth]{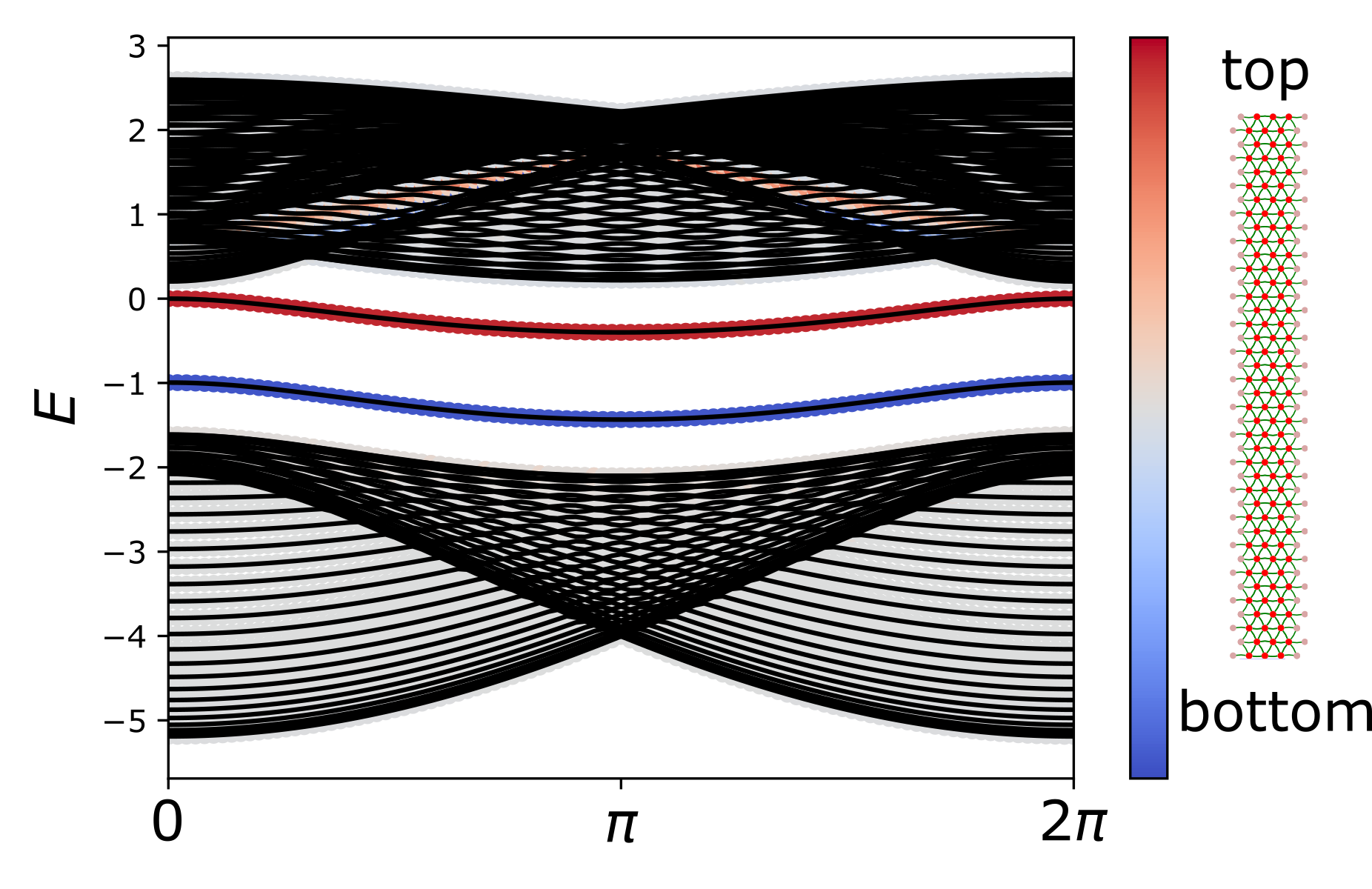}
		\caption{Ribbon spectrum of the modified triangular lattice with open boundaries on the top and bottom, showing edge modes within the bulk bandgap, using a normalization of $a=1$. The color bar shows the expectation value of the position operator in the vertical ($y$) dimension with with red (blue) denoting modes localized on the top(bottom) of the ribbon. Bulk bands appear black, being fully delocalized.}
		\label{fig3}
	\end{figure}
	
	The symmetry indicators show that the Hamiltonian Eq. \eqref{H} is that of a nontrivial phase protected by $C_3$ rotation, but it does not guarantee the existence of edge states for all finite edges. Namely, the non-zero value of the $\chi^{(3)}$ indicator here denotes a protected fractional charge per unit cell which can exist along suitably chosen boundaries, rather than the existence of edge states pinned within the bulk band gap (see also Sec. \ref{sec:wilson}) Finite boundaries that break the straight-line edge geometry will not support nontrivial edge states. This is important as the existence of the edge states is therefore gauge-dependent, being removable by a change in coordinate system or redefinition of the finite boundary, similar to those seen in Kagome lattices \cite{ni_observation_2019}. Fig. \ref{fig3} shows the spectrum of a finite ribbon of the triangular lattice with $\delta = 0.4$, a boundary that maintains the required symmetry along the top and bottom, and open boundary conditions. We see two edge modes appearing within the bulk bandgap, caused by the non-zero topological invariant $\chi^{(3)}$. The modes are pinned to the top and bottom of the ribbon. The two edge states are here shown at differing energies, which is a natural result of the edge termination being different (i.e., the unit cell is not $C_2$-symmetric, so the top and bottom edge must necessarily have differences in the hopping texture).
	
	We note here that at $\delta = 0$ there can be no edge states at any energy, but for finite $\delta \ne 0$ they will emerge from the bulk spectrum, including the Dirac-like $\delta < 0$ case. In such cases, as well as the nontrivial $\delta >0$ case studied here for small $\delta$, the edge states exist within a continuum of bulk states. Only when the finite dispersion of the nontrivial $\delta>0$ case permits a complete bandgap (here for $\delta = 0.16$) will fully isolated states \footnote{The states themselves can be removed by a surface perturbation, forced back into the bulk, as they are not required by symmetry to exist at a fixed energy in the bandgap. However, they will emerge under the circumstances presented in the model} bound to the edges appear from the continuum (see Appendix \ref{app:fin}). Fig. \ref{fig4} gives an energy diagram as a function of $\delta$ for a finite lattice, showing these isolated modes appearing for $\delta >0$. 
	
	\begin{figure}
		\centering
		\includegraphics[width = \linewidth]{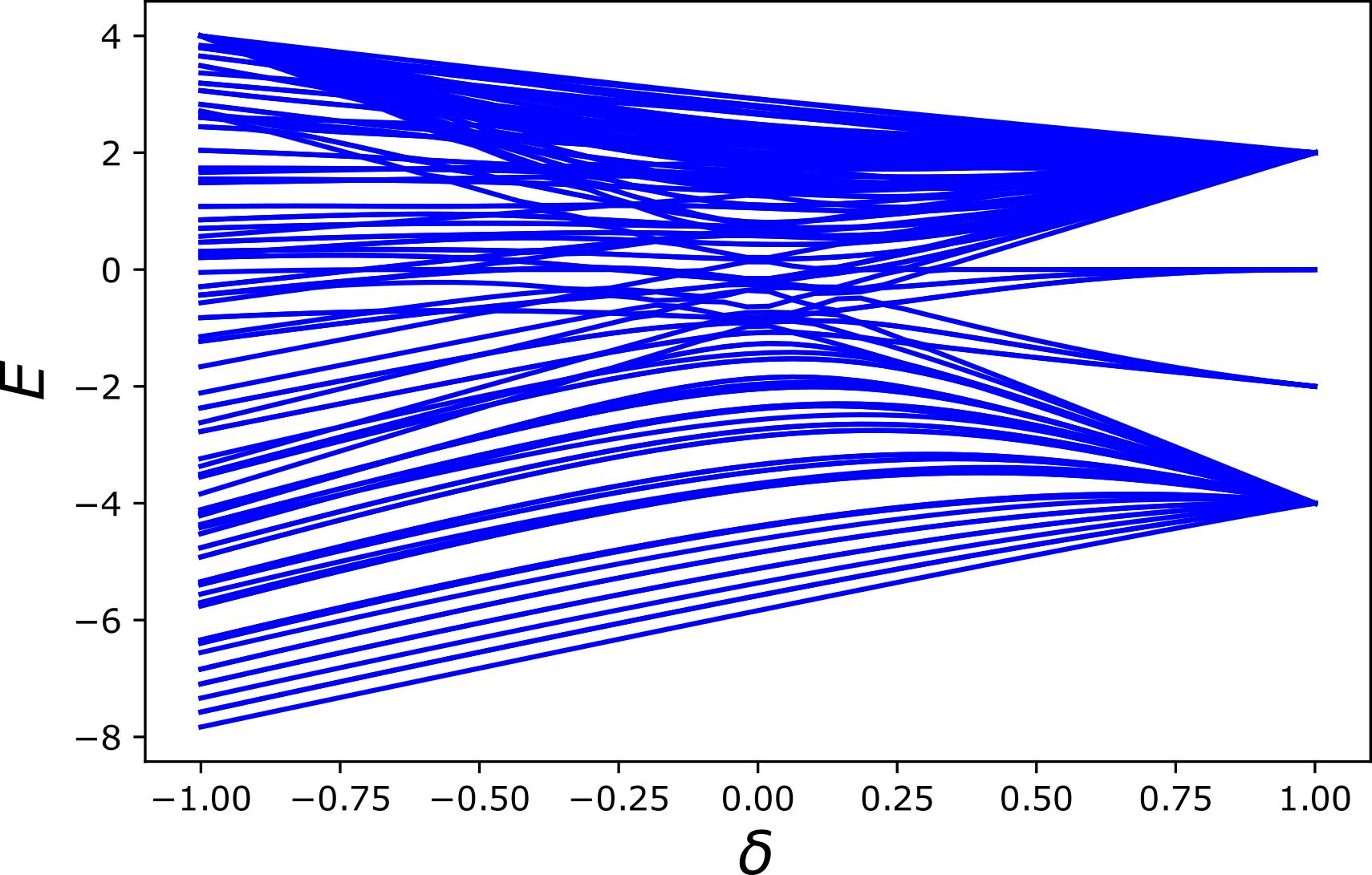}
		\caption{Edge mode existence as a parameter of the difference in hopping strengths $\delta$ for a finite triangular lattice. Pairs of modes emerge from the bulk band once the bandgap is opened large enough.}
		\label{fig4}
	\end{figure}
	\section{Symmetry Indicators for Bosonic Implementations}
	
	\subsection{Numerical Example in Photonics}\label{pho}
	A key benefit of the symmetry indicator methods used here is that they are readily applied to other physical systems via simulation. This is detailed in \cite{zhou_-chip_2021}, where a surface acoustic wave platform results in the same indicators for phonons. As a further demonstration of this, here we show the results for a 2D photonic crystal model, similar to those studied in \cite{bisharat_robust_2021} and \cite{yang_evolution_2021}. 
	
	As a computational aside, to transfer the idea of symmetry indicators to such a platform where the wave function is defined continuously over the simulation domain, the definitions for the symmetry indicators must be suitably altered. Namely, to compute the various rotational eigenvalues, the procedure can be simply performed via a scalar multiplication of the 2D eigenfield by the relevant complex number corresponding to the rotational eigenvalue. This results in another 2D eigenfield, and the value of the indicator element becomes a sum of each of these fields that match the original eigenfield. For additional details and a walkthrough of this process, see Appendix \ref{contin}. 
	
\begin{figure}
		\centering
		\includegraphics[width = \linewidth]{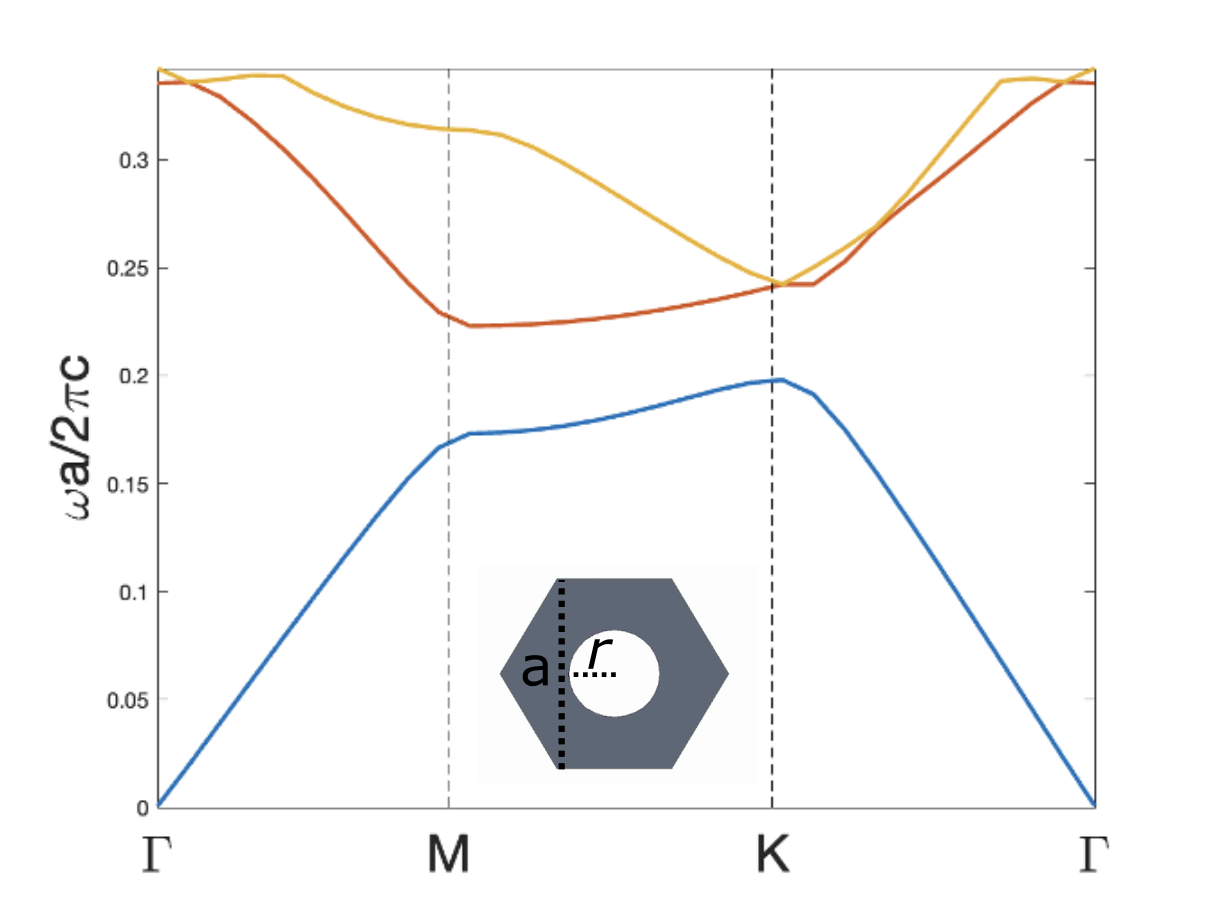}
		\caption{Photonic band structure for the first 3 bands of a 2D silicon ($\epsilon_r =11.9$) photonic crystal in a triangular lattice. Inset is the structure, where here $a = 20$ mm, $r = 4.6$ mm. The gray indicates silicon, while the white is air.}
		\label{fig:bands}
	\end{figure}

\begin{figure*}
		\centering
		\includegraphics[width = \linewidth]{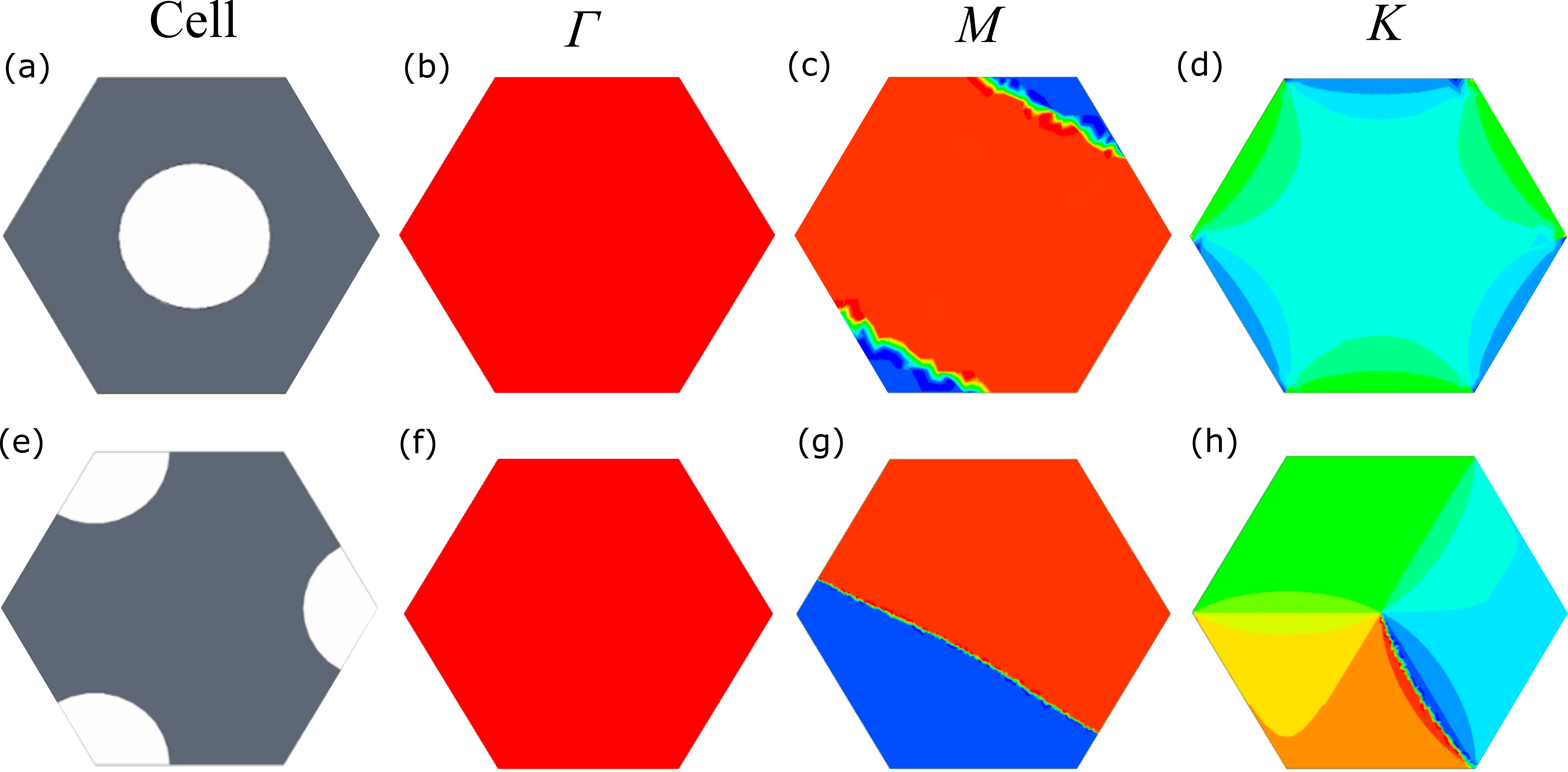}
		\caption{Computation of symmetry indicators from 2D photonic crystal phase plots. For the two definitions of unit cell (a), (e), the right three columns plot the 2D phase of the $H_z$ eigenfield at the indicated HSPs.}
		\label{fig:cells}
	\end{figure*}

	For a 2D photonic crystal in a triangular lattice composed of circular holes (for TE modes) or rods (for TM modes), we may define the unit cell first by placing the circular hole/rod at the center of a hexagon, as show in the inset to Fig. \ref{fig:bands}, which shows a representative photonic band structure calculation; note the high degree of similarity to the nontrivial case analyzed in the main text. The simulation results are done using Ansys HFSS FEM solver using a unit cell size of $a= 20$ mm, air hole radius $r=4.6$ mm, and a thin ($\ll \lambda$) height of $h=0.2$ mm, with the background material being silicon ($\epsilon_r = 11.9$). By symmetry, an equally valid choice of unit cell is one where the hole/rod is shifted, which will keep the bandstructure visually unaltered. Such a choice along with the original symmetrical choice is illustrated in Fig. \ref{fig:cells}.
	
	For the first case, the unit cell has point group $C_{6v}$, which by symmetry constraints on the Berry curvature is trivial. Likewise, under this orientation the 2D charge polarization (see Section \ref{sec:wilson}) is trivial. To compute the symmetry indicator, Fig. \ref{fig:cells} shows the $H_z$ phase profile at $\Gamma$, $M$, and $K$, which are $C_6$, $C_2$, and $C_3$-rotationally symmetric, respectively. From these, we compute the symmetry indicators (which can intuitively be seen by visualizing the rotation of the phase plot) and find $\chi^{(6)} = (0,0)$, giving a trivial phase. Hence, from this unit cell definition we do not expect any nontrivial behavior, analogous to a pure triangular lattice on a single site basis, with the difference of the existence of a bandgap. 
	
	For the second case, we shift the unit cell center, thereby placing the phase vortex observed from the edges to the center. This shift results in the point group reducing to $C_{3v}$, and we find the resulting symmetry indicator to be $\chi^{(3)} = (-1,+1)$, matching that found for the nontrivial case of the TB model. Performing a $C_2$ rotation on the unit cell results in $\chi^{(3)} = (-1,0)$, again matching the TB case. 
	
	\subsection{Connections to Recent Experiments}\label{experiments}
	The TB Hamiltonian Eq. \eqref{H} describes an idealized spinless particle on a triangular lattice, where topological bandgaps can be induced by tuning of the hopping amplitudes. Despite this idealization, the phenomenon of greatest relevance to experiments is the influence of the crystalline symmetry upon surface states. From the previous section, it can be seen that analogous surface states can be introduced into a photonic system, where the symmetry of the array of holes results in edge states along suitable boundaries. 
	
	These numerically predicted edge states have been demonstrated experimentally in both photonic \cite{bisharat_robust_2021}  and phononic \cite{zhou_-chip_2021} platforms. In these and other triangular lattice systems, the usual definition of the unit cell is the high-symmetry choice of Fig. \ref{fig:cells}(a), with crystalline $C_{6v}$ symmetry, which would naively map to the $\delta=0$ gapless case of Eq. \eqref{H}. However, as seen in the previous section, the gauge-freedom in unit cell choice permits a lower-symmetry unit cell (Fig. \ref{fig:cells}(e)), which reveals a non-trivial topology in analogy to the $C_{3v}$ models analyzed here. An important difference between Eq. \eqref{H} and such systems, however, is the atomic orbital basis used in TB models does not include the influence of the non-localized nature of classical waves \cite{lidorikis_tight-binding_1998}, which are instead faithfully represented by a basis of generalized Wannier functions \cite{albert_generalized_2000}. 
	
	Despite these differences, such a basis possesses distinct symmetry properties that match those of the atomic orbital basis employed here \cite{cloizeaux_orthogonal_1963}, and the Hamiltonian Eq. \eqref{H} yields a similar band structure to that of 2D photonic crystal realizations \cite{yang_evolution_2021}. The only major physical difference is that in the bosonic implementations a flipped copy of the lattice is used to form an interface, rather than open boundaries; as shown here such a rotation results in gauge-dependent phases, and as such also leads to edge states. Such an arrangement also provides a bandgap material on both sides of the finite edge, useful for experiments. 
	
	Furthermore, the symmetry indicator method employed here has been extended to the phononics case in a similar system \cite{zhou_-chip_2021}, and even to photonics on surface wave metallic systems \cite{wen_designing_2022}. These platforms have illustrated the high degree of robustness to perturbations of the system, including sharp angle turns and defects along the boundary. 
	
	Care must be made when applying Eq. \eqref{H} to directly model such bosonic systems for the aforementioned issues of the basis choice. Similarly, within the experimental models in Refs. \cite{bisharat_robust_2021, zhou_-chip_2021} the states manifest as propagating edge states, though the symmetry indicators used here merely protect the accumulation of edge charges. Propagating states can be expected in such experimental platforms by the inherent setting of a fixed $k$-vector by the excitation source used, coupled with the non-zero group velocity observed in their band structure. Such states can therefore be removed from the bandgap or have their propagation direction flipped by a continuous surface perturbation, and can be compared to those seen in Fig. \ref{fig3}. Nevertheless, the numerical example of the previous section shows a strong connection behind the symmetries involved and the resulting behavior of finite systems. To construct a more direct mapping between the physics of the bosonic systems and Hamiltonians on triangular lattices as studied here, it would be possible to define a triplet of orbitals on the same lattice site, which can open a bandgap without reducing the symmetry in real space \cite{wang_quantum_2016}.
	
	\section{Wilson Loop Spectra and 2D Charge Polarization Description}\label{sec:wilson}
	
	The discussion in prior sections employs the use of symmetry indicators as an efficient and general means of understanding the topology of the system, but this is not the only technique. Alternatively, the Wilson loop spectra can be used to determine the location of the Wannier centers, which gives the fractionalized charge of the lattice. This approach, like the Berry curvature, requires diagonalization of the Hamiltonian for all values within the BZ, and as such is much more computationally demanding for large systems. Unlike the curvature, however, the Wilson loop allows for another topological invariant, the charge polarization \cite{benalcazar_electric_2017}, to be computed. Here the principle is that displacement of the Wannier center from the center of the (real space) unit cell indicates a charge imbalance that is compensated by edge states on a finite sample. This section will illustrate how the Wilson loop spectra can be alternatively used to explain the behavior of the triangular lattice system studied here. We will primarily follow the preliminaries of \cite{alexandradinata_wilson-loop_2014}, which has further details for the interested reader. 
	
	First, for generic tight-binding Bloch Hamiltonian $H(\mathbf{k})$ with eigenstates $u^\mathbf{k}_n$ defined for band $n$, we first define the non-Abelian Berry connection $\mathbf{A}$ as
	\begin{equation}
		\mathbf{A}_{mn}(\mathbf{k}) \equiv i\langle u^\mathbf{k}_{m}|\nabla_\mathbf{k}|u^\mathbf{k}_n\rangle.
	\end{equation} 
	The (continuum) Wilson loop can then be described a path ordered exponential
	\begin{equation}
		\mathcal{W}(l) = T\exp\left(-i\int_l d\mathbf{l}\cdot \mathbf{A}(\mathbf{k})\right),\label{wilsonContin}
	\end{equation}
	where $l$ denotes a closed loop in reciprocal space and $T$ denotes path-ordering. The eigenvalues of Eq. \eqref{wilsonContin} encode the non-Abelian Berry phases of the Bloch bands considered. For calculation purposes, we may determine the Berry phases of a specific TB model by defining a discrete version as
	\begin{equation}
		\mathcal{\theta}(k_i)=-\text{Im}\log\prod_j\det M^{k_i,k_{j}}.\label{wilson}
	\end{equation}
	Here we have introduced an overlap matrix $M$ to handle cases of degeneracy, whose elements are defined as 
	\begin{equation}
		M_{mn}^{k_i,k_j} = \langle u_m^{k_i,k_j} | u_n^{k_i,k_{j+1}}\rangle.
	\end{equation}
	
\begin{figure}
		\centering
		\includegraphics[width=\linewidth]{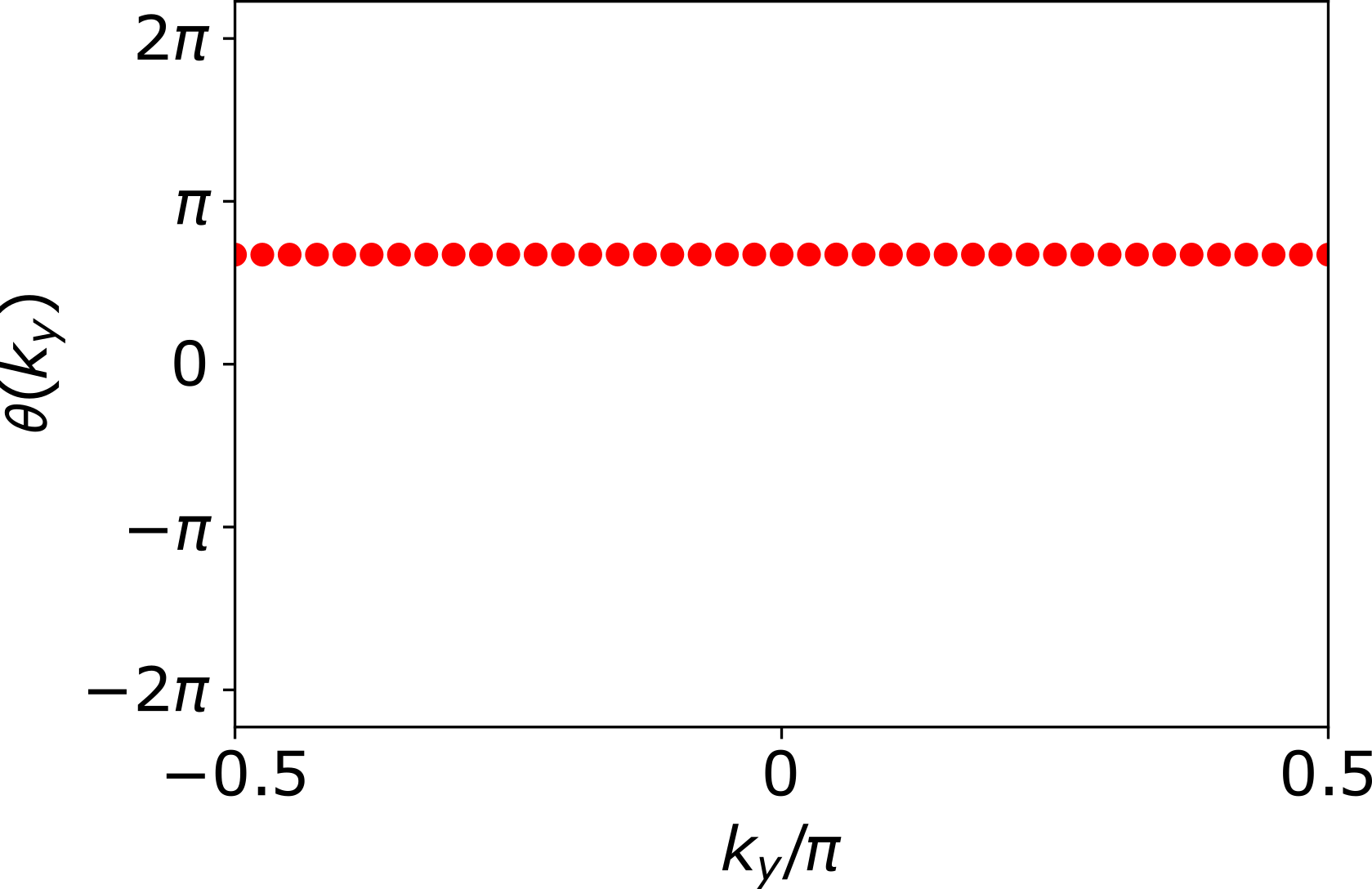}
		\caption{The Berry phase computed from the Wilson loop spectra of the $\delta>0$ TB model. It is pinned to $1/3$rd the full winding for all values across the BZ. }
		\label{fig:wilson}
	\end{figure}

	The result of computing Eq. \ref{wilson} (normalized by $2\pi$) is the location of the Wannier center for a given $k_i$. For the nontrivial $\delta > 0$ case of the triangular lattice, we find the Berry phase as shown in Fig. \ref{fig:wilson}. It is important to note that here the coordinate axes are selected such that the lattice sites are displaced symmetrically about the origin (e.g., as in Fig. \ref{fig1}(a)). The charge polarization depends on the choice of unit cell, but the location of the Wannier center with respect the physical lattice does not (see Appendix \ref{gauge} for more details). 
	
	We can see that the first band is pinned to $+1/3$ a full BZ winding. As we are concerned with the first bandgap, the important behavior is contained in the fractionalized nature of the first band alone. This indicates a nontrivial topology, which we may formalize via the 2D charge polarization given by
	\begin{eqnarray}
		P=\frac{1}{2\pi L_{\mathcal{W}}}\int_{L_{W}}Wdl
	\end{eqnarray}
	
	Here we have defined the polarization normalized by the electric charge $e$, and, based on symmetry constraints, reduced the 2D polarization to a single term (as it is equal in both directions for our case). In what follows, we have chosen the lattice vectors $\mathbf{a_1}=\mathbf{\hat{x}}$, $\mathbf{a_2}=\frac{1}{2}\mathbf{\hat{x}}+\frac{\sqrt{3}}{2}\mathbf{\hat{y}}$. From this, we find the nontrivial band to have charge of $+1/3$ for the configuration in Fig. \ref{fig1}(a) (e.g., corresponding to $\chi^{(3)}=(-1,+1)$). For the $C_2$-rotated version (corresponding to $\chi^{(3)}=(-1,0)$), we instead get $-1/3$. 
	
	As expected, we recover the same topological protection as that found via the symmetry indicators, albeit with greater computational expense. However, the usefulness of the indicators extends still further, as there is a strong connection between the symmetry indicators found and the bulk charge polarization itself. Namely, for $\chi^{(3)}$, we may consider the polarization as given by \cite{benalcazar_quantization_2019}
	\begin{equation}
		P^{(3)} = \frac{2}{3}([K_1^{(3)}]+2[K_2^{(3)}]).\label{pol}
	\end{equation}
	Noting again that we have dropped the vector component here (as both elements will be equal), and that we are still defining the polarization normalized to $e$, we can now quickly compute the polarization for our model without the full BZ information used in the Wilson loop approach. We find a polarization of $+1/3$ for the $\chi^{(3)}=(-1,+1)$ case, and $-1/3$ (or, equivalently, $+2/3$) for the $\chi^{(3)}=(-1,0)$ case. 
	
	Such a connection to the charge polarization makes the existence of the edge states seen in Fig. \ref{fig3} clearer: the states arise due to the fractionalized charge per unit cell, which are pinned to specific locations within the unit cell. When the boundary is chosen such that the charges align, a surface state can appear, but can be removed by a surface deformation. 
	
	Lastly, we note that higher order states can be induced within the triangular lattice model presented here using suitable modifications to the hopping structure, in an analogous fashion to those seen in Kagome lattices. This can be predicted by the corner charge (normalized by $e$) \cite{benalcazar_quantization_2019},
	\begin{equation}
		Q^{(3)}_{corner} = \frac{1}{3}[K_2^{(3)}]\mod 1,
	\end{equation}
	which is equal to $+1/3$ and $0$ for the $\chi^{(3)}=(-1,+1)$ and $\chi^{(3)}=(-1,0)$ cases, respectively. As the bulk charge polarization of Eq. \eqref{H} is non-zero for the gapped phase there can be no fractionalized corner charges, but the Hamiltonian can nonetheless can be combined with other crystalline models that cancel the polarization (the so-called \textit{nominal corner charges}). Under such a combination, we expect to see localized corner states appearing for finite lattices with the arrangement shown in Fig 1(a) of the main text, but not its $C_2$ rotated copy. This can be understood by considering the edge geometry of a finite lattice with $C_3$ symmetry, which naturally leads to the Wannier centers appearing on the corners for only one orientation of the unit cell (see the supplementary info of \cite{benalcazar_quantization_2019} for further details). 
	
	\section{Conclusions}
	We have demonstrated that a triangular lattice chosen with a 3-site basis and a specific hopping texture is topologically nontrivial, and can support states bound to finite edges that maintain a straight line termination. The model does not possess nonzero valley or inversion-symmetry topological invariants, and is instead described by a symmetry indicator arising from rotational eigenvalues. Our model deepens our understanding of a number of recent experimental demonstrations related to anisotropic wave/energy propagation and verifies their real space topological origin.
	
		\begin{acknowledgments}
		R.J.D. would like to thank V. Khurana for stimulating discussions. This work was supported by AFOSR grant FA9550-16-1-0093. 
	\end{acknowledgments}
	\appendix
	\section{Group Theoretic Constraints for Ideal Triangular Lattices}\label{group}
	Here we give a brief derivation of the constraints on the triangular lattice states based on group theory, much of which can be found in \cite{malterre_symmetry_2011}. 
	
	If we consider a Hamiltonian on a triangular lattice (Eq. \eqref{H} of the main text), the space group is P6mm. To understand the allowable states for the periodic case, we can impose a potential that retains the 6-fold rotational symmetry, as in the case of $\delta = 0$ of the main text. In such cases we can then analyze the behavior of the little group of the wave vector at the various high symmetry points (HSPs) \cite{tinkham_group_2003}. 
	
	At the $\Gamma$ point, the little group coincides with the point group, which is $C_{6v}$. This group contains 1D and 2D irreducible representations (irreps), and as such implies at $\Gamma$ we expect isolated as well as doubly degenerate modes. Likewise, at $K$ the little group is $C_{3v}$, which has both 1D and 2D irreps. Conversely, at the $M$ point the little group becomes $C_{2v}$, which only contains 1D irreps, and therefore any degeneracy is not required by symmetry. 
	
	From the above, we may conclude that the degeneracy along the $M-K$ path shown in the main text is not protected by symmetry, and may be broken by considering differing models of the potential. Indeed, employing perturbation theory to the nearly-free electron model with finite potential will separate the bands at the $M$ point \cite{malterre_symmetry_2011}. Nevertheless, a key aspect to the symmetry indicator method used here is that the resulting topological invariant is maintained for these alternative models, as it only requires eigenvectors at HSPs and the preservation of rotational symmetry.
	
	\section{Gauge Choices for Symmetry Indicator Methods}\label{gauge}
	For the application of the symmetry indicators from e.g., \cite{benalcazar_quantization_2019}, it is worth digressing on the importance of the gauge condition required, particularly for simple TB models like those employed here. Namely, there are two characteristics required to compute the values, the first being the generalized symmetry constraint
	\begin{equation}
		\hat{r}_nh(\mathbf{k})\hat{r}_n^\dagger = h(R_n \mathbf{k})\label{sym}
	\end{equation}
	where $h$ is the (Bloch) Hamiltonian, $\hat{r}_n$ is the desired $n$-fold rotational operator, and $R_n$ is the corresponding 2D rotation matrix acting on the crystal momentum $\mathbf{k}$. The second constraint is that placed by the HSPs $\mathbf{\Pi}$ for which the relation
	\begin{equation}
		R_n \mathbf{\Pi} = \mathbf{\Pi}\label{rot}
	\end{equation}
	holds within the periodic BZ. The combination of the above two conditions can be combined to show that, in order for the symmetry indicators to be defined, the rotational operator must commute with the Hamiltonian, $[\hat{r}_n,h]=0$. 
	
	The above conditions are innocuous enough, but there is some subtly with respect to "gauge choices," which can result in unexpected or erroneous conclusions. By "gauge," here we mean both with regards to the gauge choice of the Hamiltonian in the Bloch basis, as well as the "physical" gauge of the real-space Hamiltonian. 
	
	The first issue, the Hamiltonian's gauge choice, is seldom discussed, but has genuine consequences, particularly in tight-binding models \cite{bena_remarks_2009}. When defining a TB model, there are two main methods, the so-called "periodic gauge," wherein the wavefunction is expanded as a sum of Bloch functions, each containing their own phases,
	
	\begin{equation}
		\psi_{\mathbf{k}}=\frac{1}{\sqrt{N}}\sum_{\mathbf{R},j}c_j(\mathbf{k})e^{\mathbf{k}\cdot(\mathbf{R}+\mathbf{a_j})}|\phi_{\mathbf{R},j}\rangle,
	\end{equation}
	where $\mathbf{a_j}$ denotes the atomic location of orbital site $j$, and the "Bloch" gauge, where the all atomic sites are considered together with a single phase,
	\begin{equation}
		\tilde\psi_{\mathbf{k}}=\frac{1}{\sqrt{N}}\sum_{\mathbf{R},j}\tilde c_j(\mathbf{k})e^{\mathbf{k}\cdot\mathbf{R}}|\phi_{\mathbf{R},j}\rangle.
	\end{equation}
	
	The "Bloch" choice is the one most familiar from textbook examples, as it is both simpler to write down (hopping terms within the unit cell are real numbers) and has the benefit of being periodic in the BZ, $h(\mathbf{k}+G) = h(\mathbf{k})$ for reciprocal lattice vector $G$. However, the "periodic" choice is often more physical with respect to features like the Berry curvature \cite{dobardzic_generalized_2015} (in the main text this gauge was used for Figs. \ref{fig2}-\ref{fig:berry} for this reason). It is likewise often more natural for calculations involving electrical polarization, as in the Wilson loop spectra. The conventions are related by a unitary transformation, but there are added consequences depending on what further calculations are desired. 
	
	For the purposes of symmetry indicators, the Bloch gauge is necessary, as condition \eqref{rot} cannot be met without the periodicity of the wavefunctions. As different numerical software packages for creating TB models differ in their gauge choice, the user may arrive at incorrect answers if the wrong gauge is chosen. This can also lead to great confusion since the initial symmetry constraint \eqref{sym} will hold regardless of gauge choice, as will all physical observables. 
	
	The second issue, the "physical gauge" is more easily understood pictographically, but is no less important for the proper investigation of a given model. By "physical," we mean the coordinate system chosen in real space, and the resulting arrangement of the atomic sites. This is often not an issue for most studies, but in cases like the triangular lattice studied here, there can be a great difference between two otherwise identical models. 
	
\begin{figure}
		\centering
		\includegraphics[scale=0.75]{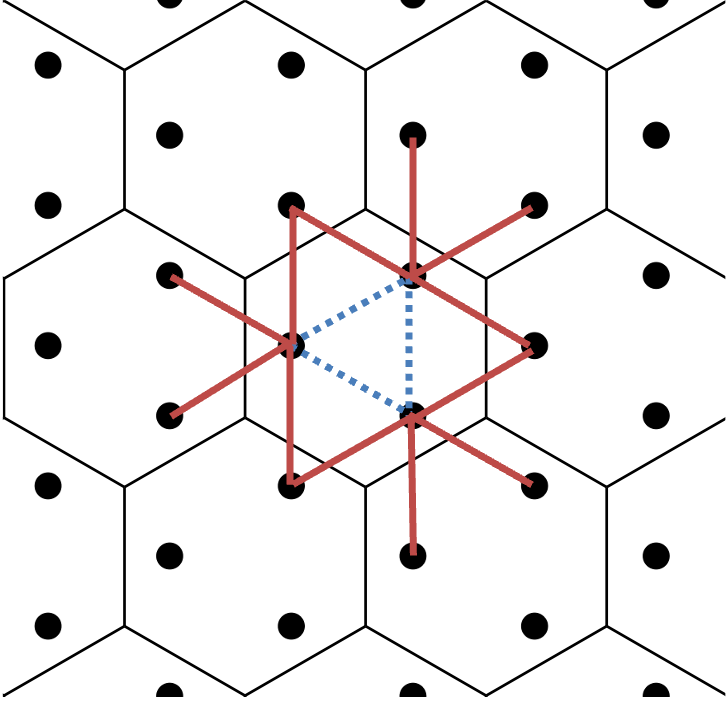}
		\caption{Alternative choice of unit cell, with identical eigenspectra but differing symmetry indicators}
		\label{fig:supCell}
	\end{figure}

	For example, suppose instead of the unit cell chosen in the main text (Fig. \ref{fig1}(a)), the choice shown in Fig. \ref{fig:supCell} is made. This unit cell has (in the Bloch gauge), the Hamiltonian kernel
	\begin{equation}
		H(\delta) = \begin{pmatrix}
			0 & h_{12}^* & h_{13}^* \\
			h_{12} & 0 & h_{23}^* \\
			h_{13} & h_{23} & 0
		\end{pmatrix}.\label{AH}
	\end{equation}
	with $h_{12} = t_+ + t_- e^{ik_x} +  t_- e^{i(k_x/2 +\sqrt{3}/2 k_y)}$, $h_{13} = t_+ + t_-  e^{-i(k_x/2 -\sqrt{3}/2 k_y)} +t_- e^{i(k_x/2 +\sqrt{3}/2 k_y)}$, and $h_{23}= t_+ + t_- e^{-i(k_x/2 -\sqrt{3}/2 k_y)}+ t_- e^{-i k_x}$.  
	
	This Hamiltonian, being related to the one used in the main text by a translation of the real space coordinates, has identical eigenspectra as Eq. \eqref{H}. However, a significant difference distinguishes them: Eq. \eqref{AH} has fully trivial symmetry indicators for all $\delta$. 
	
	This can be understood by referring the the Wilson loop spectra as analyzed in Sec. \ref{sec:wilson}. As the Wilson loop spectra gives the charge polarization, we can see that the center of charge for this model resides at the midpoint between the three lattice sites linked by blue bonds. For the model considered in the main text, this results in the charge polarization being split between three locations on the outer edge of the unit cell, while in the case of Eq. \eqref{AH}, it is symmetrically located at the center. From Eq. \eqref{pol}, we can conclude that this immediately gives trivial symmetry indicator values. This latter gauge dependency was also exploited recently for waveguiding applications \cite{wen_designing_2022}.
	
	The above discussion illustrates that, while powerful, symmetry indicators are sensitive to gauge/unit cell decisions, and can cause issues when care is not made in their use. It is also worth stressing here that this gauge-dependence may seem to contradict the notion of "topological protection" in the traditional sense of the quantum Hall effect. Systems as discussed here differ from these others in a number of ways (particularly on the termination structure), but the key feature that permits the "topological" nomenclature here is that they are still a phase defined by a \textit{global} behavior, and can be characterized by a discretely changing parameter linked to the existence of edge states. 
	
	\section{Computational Aspects of Symmetry Indicators for Tight-Binding Models}\label{comp}
	
	The symmetry indicator method employed here has gained in popularity recently, but is often difficult to follow how authors make use of it, and there are few resources to assist those who wish to perform the calculations themselves. Various rigorous arguments and proofs for these methods can be found in e.g., \cite{benalcazar_quantization_2019}. This technique is computationally very efficient, as it does not require diagonalization at all $k$ points (like the Chern number), and less mathematically involved than methods like the $\mathbb{Z}_2$ invariant. This section aims to fill in the more numerical details involved in such calculations, and hopefully make clear what is being presented.
	
	To begin with, for tight-binding models the first step is to calculate the eigenvectors of the Hamiltonian directly at the relevant HSPs of the BZ. For 2D models, this merely involves the diagonalization of at most 3 matrices (e.g., $\Gamma, M, K$ for triangular and $\Gamma, X ,M$ for square lattices), giving eigenvectors $u_j(\Pi)$ for each HSP $\Pi$ and band $j$. 
	
	The second step is to then compute the expectation value of the desired rotational operator when acting on each computed eigenvector. Construction of such operators is simple within the TB formalism, and are merely matrices that permute the given orbital sites. The eigenvalues of each rotational operator are always given as 
	\begin{equation}
		\Pi^{(n)}_p = e^{2\pi i (p-1)/n}, \quad p = 1,2,3,\cdots, n
	\end{equation}
	for an $n$-fold rotational operator. We therefore know in advance that the computation of the given symmetry indicators will involve counting up these values, and any computation that differs from them is likely an error (commonly noticed due to improper handling of degeneracies, which will be covered shortly, or a gauge error, which will be considered in Appendix \ref{gauge}). 
	
	Computing the expectation for a single isolated band involves a simple inner product, reproduced from the main text as
	\begin{equation}
		\langle \hat{r}^u_n(\Pi)\rangle=\langle u(\Pi) | \hat{r}_n | u(\Pi) \rangle.\label{exp}
	\end{equation}
	In the above, $u(\Pi)$ denotes the eigenvector computed in the first step for the HSP $\Pi$, and $\hat{r}_n$ is the matrix representation of the rotational operator of order $n$, with the size of $\hat{r}_n$ being determined by the number of basis elements of the vectors. 
	
	The above equation will give one of the eigenvalues of $\hat{r}_n$, which may then be used for the later steps in computing the invariant. However, it is often the case, especially for more complex bandstructures, that degeneracies occur the the HSPs in question. As mentioned in the main text, the resolution to this is to consider the overlap matrix formulation of Eq. \ref{exp}, given as 
	
	\begin{widetext}
	\begin{equation}
		S(\Pi) = \begin{pmatrix}
			\langle u_1(\Pi) | \hat{r}_n | u_2(\Pi) \rangle & \langle u_1(\Pi) | \hat{r}_n | u_3(\Pi) \rangle & \cdots & \langle u_1(\Pi) | \hat{r}_n | u_M(\Pi) \rangle \\
			\langle u_2(\Pi) | \hat{r}_n | u_1(\Pi) \rangle & \ddots & &\vdots\\
			\vdots\\
			\langle u_M(\Pi) | \hat{r}_n | u_1(\Pi) \rangle & \cdots & &  \langle u_M(\Pi) | \hat{r}_n | u_M(\Pi) \rangle 
		\end{pmatrix}
	\end{equation}
\end{widetext}
	for a given manifold of $M$ degenerate bands at HSP $\Pi$. The eigenvalues of this matrix provide the desired expectation values of the rotational operator. 
	
	Once the expectation values are computed, the final step is to count the number of each eigenvalue and subtract the number located at $\Gamma$, written in general as
	\begin{equation}\label{eq:op}
		[\Pi^{(n)}_p] = \#\Pi^{(n)}_p - \#\Gamma^{(n)}_p
	\end{equation}
	This final step is less clear notationally, as indicated by the use of the $\#$ sign to mean "count the number of." The above is merely stating that to find the integer valued invariant element $[\Pi^{(n)}_p]$ for an $n$-fold rotation at HSP $\Pi$ that has eigenvalue $\Pi^{(n)}_p$, we have to count the number of bands with that same eigenvalue at $\Pi$, count the number of bands with that eigenvalue at $\Gamma$, and subtract the two counts. Note that in doing so we are forced to decide where to set the Fermi level (or, to extend the discussion to bosonics, the desired frequency), which determines the number of bands we must count the eigenvalues for. 
	
	Each calculation of the above results in a single integer. Such integers alone do not constitute the topological invariant per se, here called $\chi^{(n)}$, but rater are the elements thereof. The previous step can be done for any allowed eigenvalue and rotational operator, but, as shown at length in \cite{benalcazar_quantization_2019}, the total number of distinct combinations that are needed to properly define $\chi^{(n)}$ is much smaller. Specifically, we may write the required values as \cite{benalcazar_quantization_2019}
	\begin{align*}
		\chi^{(4)} &= ([X_1^{(2)}],[M_1^{(4)}],[M_2^{(4)}])\\
		\chi^{(2)} &= ([X_1^{(2)}],[Y_1^{(2)}],[M_1^{(2)}])\\
		\chi^{(6)} &= ([M^{(2)}_1],[K^{(3)}_1])\\
		\chi^{(3)} &= ([K^{(3)}_1],[K^{(3)}_2]).
	\end{align*}
	The above can describe all $n$-fold rotationally symmetric 2D systems for the allowed $n = 2,3,4,6$. By repeating the above steps, the invariant $\chi^{(n)}$ may be computed efficiently for any Hamiltonian. 
	
	The above walkthrough provides the "how" of computing rotational invariants, but does not directly provide insight into what is physically causing the topological distinction. An intuitive means of understanding what a nonzero $\chi^{(n)}$ is to consider a single isolated band (as was done in the main text). In such a case, the expression for each element of $\chi^{(n)}$ reduces to a single yes-no question on whether the band has the eigenvalue in question, and comparing that to the same question at $\Gamma$. For the element to be nontrivial, there necessarily must be a difference between $\Gamma$ and the chosen HSP. More concretely, the rotational behavior of the eigenvector \textit{must} change as it passes from $\Gamma$ to a given HSP. 
	
	This, then, gives the "topological" aspect: since a symmetry property changes for continuously defined bands at differing HSPs, the introduction of a finite edge (or other suitable termination) results in states that are trapped on that same edge, analogous to the edge states of other topological effects. 
	
		\section{Surface States in the Dirac ($\delta<0$) Case}\label{app:fin}
\begin{figure}
		\centering
		\includegraphics[width = \linewidth]{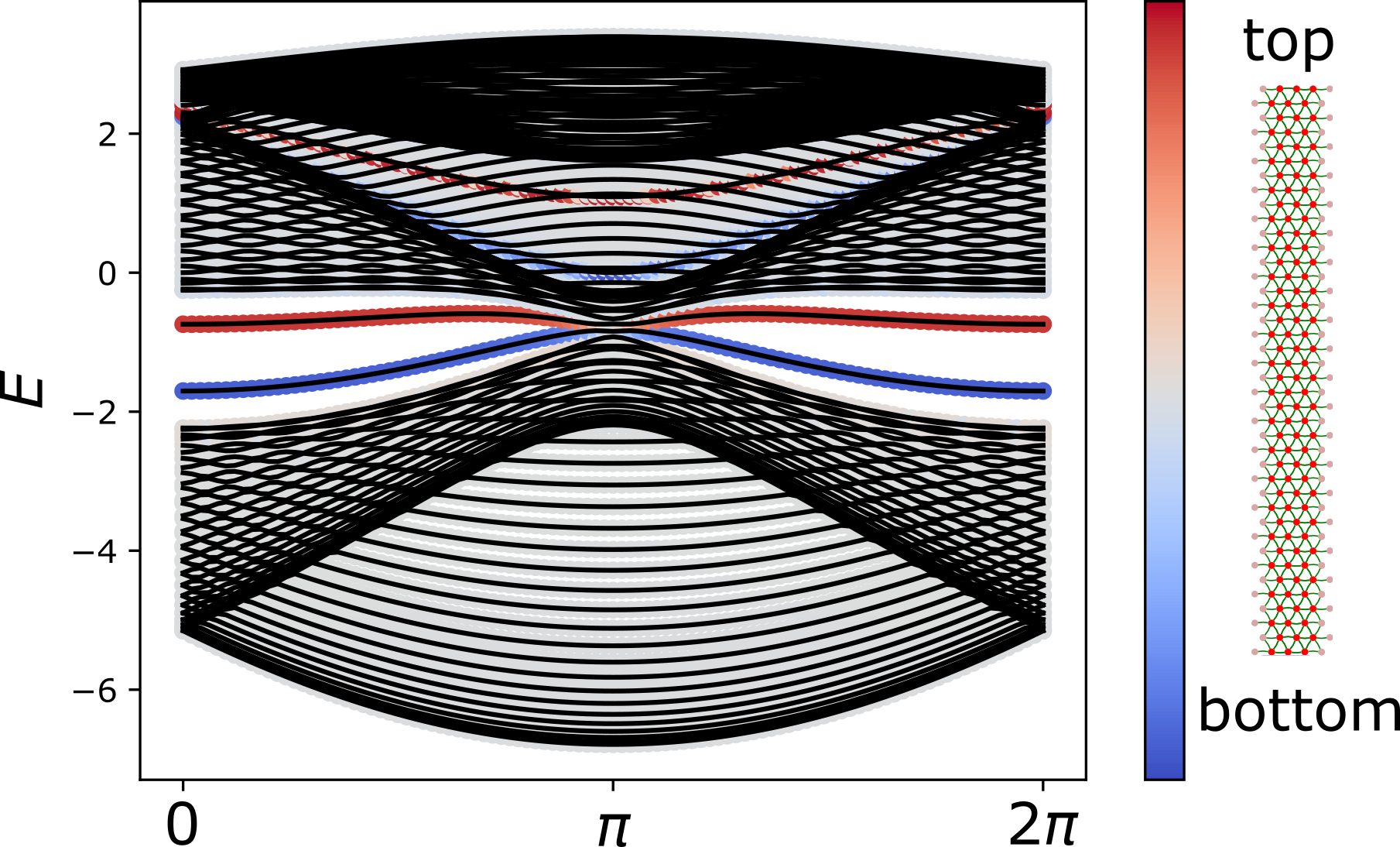}
		\caption{Ribbon spectrum of the modified triangular lattice with open boundaries on the top and bottom for the Dirac ($\delta<0$) case, showing edge states, using a normalization of $a=1$. The color bar shows the expectation value of the position operator in the vertical ($y$) dimension with with red (blue) denoting modes localized on the top(bottom) of the ribbon. Bulk bands appear black, being fully delocalized.}
		\label{fig:diracRib}
	\end{figure}

	In the case of the gapless Dirac case, with $\delta<0$, there is no bandgap, but nevertheless states exist that are localized at the edges of the system. Without a bandgap they naturally coexist with bulk states, which is shown via a ribbon spectra for a semi-infinite model shown in Fig. \ref{fig:diracRib}. As discussed in Sec. \ref{sec:fin}, as $\delta$ is tuned to the transition point of $\delta=0$, a bulk bandgap is opened, which permits these localized states to be isolated from the continuum under the parameter regime given in Fig. \ref{fig4}. Such states cannot be naively classified as topological (as there is no bandgap), but are still related to the symmetry-enforced existence of Dirac cones (see Fig. \ref{fig1}(d)). 
	
	\section{Computational Aspects of Symmetry Indicators for Continuously Defined Models}\label{contin}
	
	Sec. \ref{pho} shows a calculation of the symmetry indicators for a photonic crystal model that displays similar behavior to the TB model studied in the paper. However, such a calculation differs from the discrete TB model as explained in Appendix \ref{comp}, since systems like photonic or phononic crystals have eigenfunctions that are continuously defined across their unit cell area, and are therefore represented numerically by $N\times M$ matrices, rather than vectors. The physics is fundamentally the same, but the numerical details must be suitably adjusted to handle this. 
	
	In short, the method employed to determine $\chi^{(3)}$ for a continuously defined model is as follows:
	\begin{enumerate}
		\item Numerically solve the eigenvalue problem for the unit cell at the $K$ and $\Gamma$ HSPs and extract the phase profile over the full real space unit cell boundaries.
		\item Multiply these two phase profiles, point by point, by the numerical value corresponding to the 3-fold operator to be considered(e.g., $1,e^{\pm2\pi i/3}$). This results in three altered profiles for each HSP (6 in total, though two will just be the original,  unaltered profiles corresponding to the eigenvalue 1).
		\item Compare each of these altered phase profiles to that of the original phase profile rotated by 120 degrees. This is easily done visually, or can be automated via a point-wise comparison. Whichever altered profile matches is the correct eigenvalue corresponding to that operator acting on that HSP. 
		\item Apply Eq. \eqref{eq:op} for all modes up to the desired bandgap/eigenfrequency to retrieve the elements for $\chi^{(3)}$. 
	\end{enumerate}
	
	This process can be readily adapted to any other rotational operator, and is essentially a point-by-point version of Eq. \eqref{exp}, amenable to automated numerical computations. 
	
	To see how the above procedure is equivalent to Eq. \eqref{exp} mathematically, we can consider that the physical effect of rotation operators  $R_n$ is to rotate the locations in 2D space of lattice sites. If instead of a vector of basis sites we have a continuously defined eigenfunction of 2 dimensions $|\psi(x,y)\rangle$, the operator will act on the physical coordinates $(x,y)$. To then compute the desired expectation value, we generalize the inner product definition to the $L^2$ norm to find
	\begin{equation}
		\langle \psi(x,y)|R_\theta|\psi(x,y)\rangle=\int_\textit{cell} \psi(x,y)^\dagger R(\theta) \psi(x,y) dxdy= r_\theta,
	\end{equation}
	where $R(\theta)$ is the rotation matrix and $r_\theta$ are its eigenvalues. This definition is not immediately useful in the case of numerically computed eigenfunctions, where instead we have a discretely defined matrix of complex field values $\psi_{nm}$ up to a given resolution $\delta r$. We may instead construct a matrix $R_{\theta}$ that performs the rotation on each eigenfield value to enact the rotation numerically, and compute the inner product discretely as 
		\begin{equation}
		\langle \psi_{nm}|R_\theta|\psi_{nm}\rangle=\sum_\textit{nm} \psi_{nm}^\dagger R_\theta \psi_{nm}= r_\theta
	\end{equation}
	The above is formally equivalent to Eq. \ref{exp} in the limit of $\delta r\rightarrow 0$, under the same symmetry constraints Eqs. \eqref{sym}-\eqref{rot} for the matrix defining the Hamiltonian. However, this definition is cumbersome to apply, as the matrix $R_\theta$ is not a simple rotation matrix. The procedure outlined above is essentially working in reverse of this, where we assume the eigenvalue, apply it to the field, then rotate the field visually to compare it.  

		\bibliography{citesV3}
\end{document}